\documentclass[useAMS,usenatbib]{mn2e}

 \usepackage{times}

\def\pcm3{{\rm\thinspace cm^{-3}}}

\def\contcaption{\@conttrue\SFB@caption\@captype}

\def\n_h{{\rm n_{H}}}

\def\NH1{{$N_{\rm HI}~$}}

\def\ga{{\rm\thinspace gauss}}

\def\approxlt{\mathrel{\hbox{\rlap{\lower .5ex \hbox {$\sim$}}
        \raise .15 ex \hbox{$<$}}}}
\def\approxgt{\mathrel{\hbox{\rlap{\lower .5ex \hbox {$\sim$}}
        \raise .15 ex \hbox{$>$}}}}

\def\la{\mathrel{\hbox{\rlap{\hbox{\lower4pt\hbox{$\sim$}}}\hbox{$<$}}}}
\def\ga{\mathrel{\hbox{\rlap{\hbox{\lower4pt\hbox{$\sim$}}}\hbox{$>$}}}}
\newbox\grsign \setbox\grsign=\hbox{$>$} \newdimen\grdimen
\grdimen=\ht\grsign
\newbox\simlessbox \newbox\simgreatbox \newbox\simpropbox
\setbox\simgreatbox=\hbox{\raise.5ex\hbox{$>$}\llap
     {\lower.5ex\hbox{$\sim$}}}\ht1=\grdimen\dp1=0pt
\setbox\simlessbox=\hbox{\raise.5ex\hbox{$<$}\llap
     {\lower.5ex\hbox{$\sim$}}}\ht2=\grdimen\dp2=0pt
\setbox\simpropbox=\hbox{\raise.5ex\hbox{$\propto$}\llap
     {\lower.5ex\hbox{$\sim$}}}\ht2=\grdimen\dp2=0pt
\def\simgreat{\mathrel{\copy\simgreatbox}}
\def\simless{\mathrel{\copy\simlessbox}}

\title[]{A new detailed examination of white dwarfs in NGC3532 and NGC2287\thanks{Based on observations made with ESO telescopes at the La Silla Paranal Observatory under programme IDs 079.D-0490(A) and 080.D-0654(A)}}
\author[P. D. Dobbie et al.]{P. D. Dobbie,$^{1}$\thanks{E-mail:pdd@aao.gov.au} R. Napiwotzki,$^{2}$ M. R. Burleigh,$^{3}$ 
K. Williams,$^{4}$ R. Sharp,$^{1}$ M. A. Barstow,$^{3}$ \newauthor S. L. Casewell$^{3}$  and I. Hubeny$^{5}$\\ 
$^{1}$ Anglo-Australian Observatories, PO Box 296, Epping, NSW 1710, Australia \\ 
$^{2}$ Science \& Technology Research Institute, University of Hertfordshire,
College Lane, Hatfield, AL10 9AB \\ 
$^{3}$ Department of Physics and Astronomy, University of Leicester, University Road, Leicester LE1 7RH, UK\\ 
$^{4}$ Dept. of Astronomy, University of Texas at Austin, TX 78712, USA \\
$^{5}$ Steward Observatory and Department of Astronomy, University of Arizona, Tucson, AZ 85721, USA\\}

\begin{document}

\date{Accepted 2009 February 23.  Received 2009 February 10; in original form 2008 December 2}

\pagerange{\pageref{firstpage}--\pageref{lastpage}} \pubyear{2002}

\maketitle

\label{firstpage}

\begin{abstract}

We present the results of a photometric and spectroscopic study of the white dwarf candidate members of the intermediate age open 
clusters NGC3532 and NGC2287. Of the nine objects investigated, it is determined that six are probable members of the clusters, 
four in NGC3532 and two in NGC2287. For these six white dwarfs we use our estimates of their cooling times together with the cluster
ages to constrain the lifetimes and masses of their progenitor stars. We examine the location of these objects in initial mass-final
mass space and find that they now provide no evidence for substantial scatter in initial mass-final mass relation as suggested by previous 
investigations. Instead, we 
demonstrate that, when combined with current data from other solar metalicity open clusters and the Sirius binary system, they hint 
at an IFMR that is steeper in the initial mass range 3M$_{\odot}$$\simless$M$_{\rm init}$$\simless$4M$_{\odot}$ than at progenitor 
masses immediately lower and higher than this. This form is generally consistent with the predictions of stellar evolutionary models and can aid 
population synthesis models in reproducing the relatively sharp drop observed at the high mass end of the main peak in the mass distribution 
of white dwarfs.

\end{abstract}


\section{Introduction}

Standard stellar evolutionary theory predicts the existence of a positive correlation between the main sequence mass 
of a star (M$\simless$10M$_{\odot}$) and the mass of the white dwarf remnant left behind after it has expired. This is
frequently refered to as the initial mass-final mass relation (IFMR). A secure and detailed knowledge 
of the form of the IFMR is important to a number of very active areas of astrophysical research. For example, the 
relation is a key ingredient of models of the chemical evolution of the Galaxy as it provides an estimate of the amount
of gas, enriched with C, N and other metals, a low or intermediate mass star returns to the ISM (e.g. Carigi, Colin
\& Peimbert 1999).  Moreover, understanding the form of the IFMR is crucial to deciphering information locked up in the white 
dwarf luminosity functions of stellar populations (e.g. Jeffery et al. 2007, Oswalt et al. 1996). The shape of the upper end
of the IFMR is of special interest as it places limits on the minimum mass of a star that will experience a Type II supernova 
explosion. With robust constraints on this mass, for example, the observed diffuse SNe neutrino background can serve as an 
empirical normalisation check on estimates of the star formation history of the Universe (e.g. Hopkins et al. 2006). 

Unfortunately, due to the complexity of the physical processes occuring within stars, especially during the final stages of
the stellar lifecycle, the form of the IFMR is rather difficult to ascertain from first principles. For example, the mass of 
the stellar core which ultimately becomes the white dwarf is probably modified with each thermal pulse cycle on the asymptotic 
giant branch (AGB). The final remnant mass predicted by evolutionary models is therefore dependent on the length of time a 
star is assumed to spend on the AGB. However, evolution on the AGB is terminated by the removal of the stellar envelope so the
predicted duration of this phase is susceptible to the assumptions made about the rate at which envelope mass is lost (e.g. 
Iben \& Renzini 1983). While significant inroads are being made in the theoretical understanding of mass loss on the AGB (e.g.
Bowen 1988, Wachter et al. 2002), a comprehensive and robust physical treatment remains elusive.

Therefore, empirical data currently play a crucial role in improving our understanding of the form of the IFMR. Arguably the 
best way at present to obtain observational based constraints on the IFMR is through the study of the white dwarf members of 
open star clusters. The ages of these coeval populations can be determined from the location of the main-sequence turn-off 
or the lithium depletion boundary in the sequence of the lowest mass members (e.g. Pleiades; Stauffer et al. 1998). The progenitor
star lifetimes and ultimately their masses can then be estimated by calculating the difference between the age of the parent 
population and the cooling times of white dwarf members. It is worth pointing out, however, that as theoretical input is required 
at this latter stage, in the form of stellar evolutionary models, constraints acquired in this way are more appropriately termed 
semi-empirical (e.g. Weidemann \& Koester 1983). 

With greater access to 8m class telescopes and the availability of improved instrumentation, the last few years has seen a flurry 
of new studies of the white dwarf candidate members of open star clusters aimed at constraining the form of the IFMR. For example, 
Claver et al. (2001) have presented the results of a detailed photometric and spectroscopic study of five white dwarf members of Praesepe.
Williams, Bolte \& Koester (2004) have undertaken a spectroscopic study of the degenerate candidate members of NGC2168. Kalirai 
et al. (2005) have obtained multi-object spectroscopy of well over a dozen white dwarf candidate members of NGC2099. Dobbie et al. 
(2004,2006a) and Casewell et al. (2009) have identified and spectroscopically analysed a number of new white dwarfs in Praesepe. Williams \& Bolte
(2007) have investigated degenerate candidate members of both NGC6633 and NGC7063, while Kalirai et al. (2008) present spectroscopy 
of white dwarf candidate members of two older open clusters, NGC7789 and NGC6819. Most recently, Rubin et al. (2008) have unearthed six 
white dwarf candidate members of  the intermediate aged cluster NGC1039. Combined, these studies have increased the number of 
data points on the semi-empirical IFMR by over a factor of two. However, only the work of Williams, Bolte \& Koester (2004, 2008) and Rubin et al. 
(2008) has provided a substantial number of new points in the M$_{\rm init}$$\simgreat$4M$_{\odot}$ regime. The upper portion of the 
IFMR remains rather poorly constrained, with large scatter amongst the data, the bulk of which currently comes from only three clusters, NGC1039, 
NGC2516 and NGC2168. Indeed, to further complicate the situation, the metalicity of NGC2168 is subsolar by approximately a factor two, and thus 
the data points from this cluster might not be expected to occupy the same region of initial mass-final mass space as those from solar metalicity 
clusters such as the Pleiades, NGC2516 and NGC1039 (e.g. Marigo 2001).

In a bid to improve the current state-of-affairs, we have recently focused our efforts on open clusters with $\tau$$\simless$300Myrs, 
corresponding to the lifetime of a M$\approx$4M$_{\odot}$ star, and with metalicities close to solar so that the influence of this 
parameter on the data from cluster to cluster is minimal or even negligible. Here we present the results of a new investigation of the nine white dwarf
candidate members of NGC2287 (M41) and NGC3532. In the next section we review the parameters of these two clusters and the work previously 
undertaken on their white dwarf populations. Next, we describe the acquisition and analysis of new photometry and spectroscopy for these 
objects and then use these to re-assess cluster membership status. Finally, we examine those white dwarfs with parameters consistent with being 
members of the two clusters in the context of the IFMR.

\begin{figure*}
\vspace{400pt}
\includegraphics{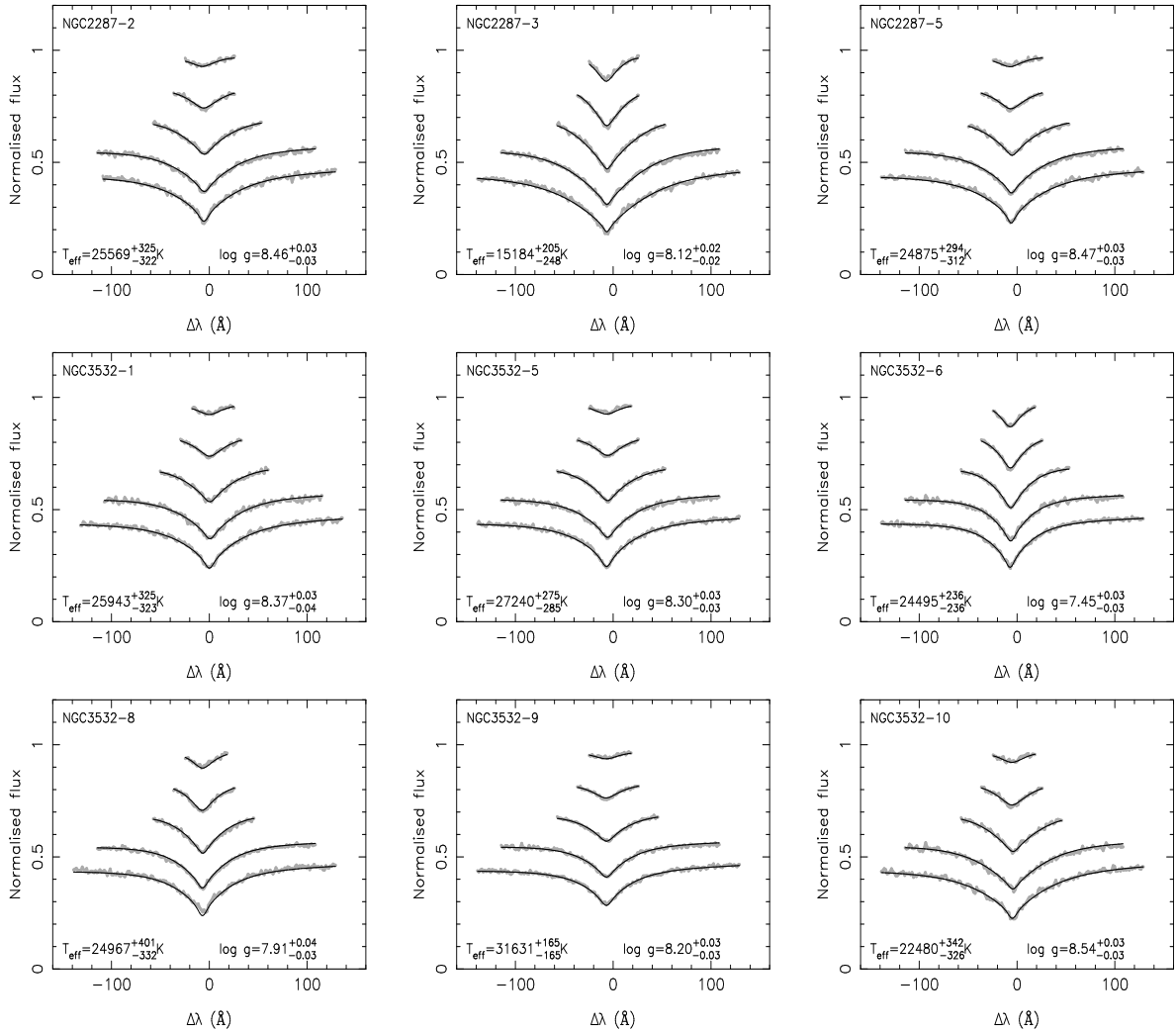}
\caption{The results of our fitting of synthetic profiles (thin black lines) to the observed Balmer lines, H-$\beta$ to H-8, of the nine white dwarf candidate 
members of NGC3532 and NGC2287 (thick grey lines). The flux$_{\lambda}$ units are arbitrary.}
\end{figure*}

\section[]{Two open clusters ripe for a new 8m study}

\subsection{White dwarf candidate members of NGC3532}

NGC3532 (11$^{h}$05$^{m}$ -58$^{\circ}$45$'$, J2000.0) is a comparatively rich and nearby open cluster yet has been relatively little studied,
 particularly in 
the last two decades. This is probably because it appears in projection against the Galactic plane (b$\sim$1.3$^{\circ}$). Despite this 
low Galactic latitude, extinction along the line of sight to the cluster is low. For example, Fernandez \& Salgado (1980) and Meynet, Mermilliod \& 
Maeder (1993) use $UBV$ photometry to determine E($B-V$)$\approx$0.04, from Str\"omgren data Eggen (1981) estimate E($b-y$)=0.023 (which equates to 
E($B-V$)$\approx$0.03) and Nicolet (1981) find E($B-V$)=0.052$\pm$0.010 using Geneva photometry. Moreover, NGC3532 appears to have a metalicity close
to the solar value. From a photometric investigation of the giant members, Claria \& Lapasset (1988) found no evidence for a significant UV excess 
or cyanogen anomaly while Twarog et al. (1997) concluded from an independent DDO based study that [Fe/H]=-0.02.

While it is unfortunate that there are less than a handful of age estimates for NGC3532 in the recent literature, at least the few that are available
appear to be consistent. For example, Meynet, Mermilliod \& Maeder (1993) compared isochrones generated from the stellar models of Schaller et al. (1992), 
which include a moderate level of convective core overshooting, and UBV photometry to obtain $\tau$=316Myrs. Koester \& Reimers (1993), using 
moderate core overshoot models, determined an age of $\tau$=302$\pm$154Myrs for a turn-off absolute magnitude of M$_{V}$=-0.75$\pm$0.25. Kharchenko
et al. (2005) estimate an age of $\tau$=282Myrs based on photometry from the 2MASS Point Source Catalogue (PSC; Skrutskie et al. 2006) and the 
stellar modelling of the Padova group (e.g. Girardi et al. 2000). Therefore, it appears reasonable to conclude that the age of NGC3532 lies 
within the range $\tau$=300$\pm$25Myrs. 

Robichon et al. (1999) used Hipparcos data to determine the cluster distance modulus to be m-M=8.04$^{+0.37}_{-0.32}$. This is notably less than the 
recent estimate of m-M=8.61 obtained from 2MASS PSC data (Kharchenko et al. 2005). Nevertheless, these two determinations bracket the results from most
other studies. Fernandez \& Salgado (1980) estimate m-M=8.45$\pm$0.27, Eggen (1981) find m-M=8.5$\pm$0.25 while Meynet, Mermilliod \& Maeder (1993) conclude
that m-M=8.35. Reimers \& Koester (1989) exploited this relative proximity and the low levels of foreground reddening, using deep UV and red Schmidt plates to identify
seven blue candidate white dwarf cluster members. Low resolution spectroscopy obtained with the 3.6m ESO telescope and the Faint Object Spectrograph and Camera 
(EFOSC) confirmed the degenerate nature of three of these stars. A subsequent extended photometric and spectroscopic survey of NGC3532 unearthed a further
three white dwarf candidate members (Koester \& Reimers 1993).

While analysis of the EFOSC spectroscopy confirmed that these white dwarfs are young ($T$$_{\rm eff}$$\approx
$20000-30000K), as would be expected if they are members of NGC3532, due to the low signal-to-noise of this data blueward of 4000\AA, the uncertainties in 
determinations of their surface gravities and hence in estimates of their masses and cooling times, remain large (e.g. see Koester \& Reimers 1993).  The
location of these white dwarfs in initial mass-final mass space as revealed by a number of recent studies e.g. Ferrario et al. (2005), Catalan et al. (2008)
is suggestive of significant dispersion in the IFMR. A large intrinsic scatter in the relation would have important implications for our understanding of 
stellar evolution and pose a huge complication for investigations which rely on the IFMR being close to monotonic (e.g. Jeffery et al. 2007). However, before
firm conclusions are drawn on this matter, better photometric and spectroscopic data should be acquired for these stars so that the cluster membership status 
of each can be soundly examined and more stringent limits placed on their masses and cooling times.

\begin{table*}
\begin{minipage}{142mm}
\begin{center}
\caption{Details of the nine white dwarf candidate members of NGC3532 and NGC2287. Masses and cooling times for 
each star have been estimated using the mixed CO core composition ``thick H-layer'' evolutionary calculations of 
the Montreal Group (e.g. Fontaine, Brassard \& Bergeron 2001). The errors in absolute magnitudes, masses and cooling 
times shown here are derived by propagating more realistic uncertainties in effective temperature and surface gravity
of 2.3\% and 0.07dex respectively.}

\label{wdmass}
\begin{tabular}{cccccccc}
\hline
 ID in lit. &$T$$_{\rm eff}$(K)$^{*}$ & log $g$$^{*}$ &  m$_{V}$  &  M$_{V}$ & M(M$_{\odot}$)  &  $\tau_{c}$ (Myrs) & M$_{\rm init}$(M$_{\odot}$) \\
\hline

 NGC2287-2 & $25569^{+325}_{-322}$ & $8.46^{+0.03}_{-0.02}$ & 20.32$\pm$0.05 & $11.06^{+0.12}_{-0.12}$ & $0.91\pm0.04$ & $81^{+15}_{-13}$  &  4.45$^{+0.58}_{-0.38}$\\ \\
 NGC2287-3 & $15184^{+205}_{-248}$ & $8.12^{+0.02}_{-0.02}$ & 19.82$\pm$0.03 & $11.31^{+0.10}_{-0.10}$ & $0.68\pm0.04$ & $229^{+27}_{-24}$ &  -\\ \\
 NGC2287-5 & $24875^{+294}_{-312}$ & $8.47^{+0.03}_{-0.03}$ & 20.54$\pm$0.03 & $11.13^{+0.12}_{-0.12}$ & $0.91\pm0.04$ & $92^{+16}_{-14}$  &  4.57$^{+0.64}_{-0.43}$\\ \\

 NGC3532-1 & $25943^{+325}_{-323}$ & $8.37^{+0.03}_{-0.04}$ & 19.16$\pm$0.02 & $10.87^{+0.12}_{-0.11}$ & $0.86\pm0.04$ & $60^{+13}_{-12}$  &  3.83$^{+0.18}_{-0.15}$\\ \\
 NGC3532-5 & $27240^{+275}_{-285}$ & $8.30^{+0.03}_{-0.03}$ & 19.01$\pm$0.02 & $10.66^{+0.11}_{-0.12}$ & $0.82\pm0.04$ & $38^{+10}_{-9}$   &  3.71$^{+0.15}_{-0.13}$\\ \\
 NGC3532-6 & $24495^{+236}_{-236}$ & $7.45^{+0.03}_{-0.03}$ & 19.28$\pm$0.02 & $ 9.56^{+0.11}_{-0.11}$ & $0.40\pm0.04$ & $17^{+1}_{-1}$    &  -\\ \\
 NGC3532-8 & $24967^{+401}_{-332}$ & $7.91^{+0.04}_{-0.03}$ & 19.17$\pm$0.02 & $10.23^{+0.11}_{-0.11}$ & $0.59\pm0.04$ & $19^{+3}_{-1}$    &  -\\ \\
 NGC3532-9 & $31631^{+138}_{-165}$ & $8.20^{+0.03}_{-0.03}$ & 18.47$\pm$0.02 & $10.18^{+0.11}_{-0.12}$ & $0.76\pm0.04$ & $10^{+3}_{-1}$    &  3.57$^{+0.12}_{-0.11}$\\ \\
 NGC3532-10& $22480^{+342}_{-326}$ & $8.54^{+0.03}_{-0.03}$ & 19.82$\pm$0.02 & $11.44^{+0.12}_{-0.12}$ & $0.96\pm0.04$ & $149^{+22}_{-19}$ & 4.58$^{+0.47}_{-0.33}$\\

\hline
\end{tabular}
\end{center}
$^{*}$ Formal fit errors - see text for further details. 
\end{minipage}
\end{table*}

\subsection{White dwarfs candidate members of NGC2287 (M41)}

An early detailed investigation of the moderately populated NGC2287 (06$^{h}$46$^{m}$ -20$^{\circ}$45$'$, J2000.0) determined the distance to the cluster to be D=725pc and levels of extinction along this line of sight to be very low, E($B-V$)=0.01 (Cox 1954). These estimates are corroborated by the results of more recent studies where state-of-the-art theoretical isochrones have been fit to the observed cluster sequence. For example, based on an analysis of $UBV$ photometry of cluster members Meynet, Mermilliod \& Maeder (1993) derive a distance modulus of m-M=9.15 and estimate E($B-V$)=0.01. Kharchenko et al. (2005) determine m-M=9.30 and E(B-V)=0.03 based on a study of near-IR data from the 2MASS PSC (Skrutskie et al. 2006). Sharma et al. (2006) estimate m-M=9.26 and E(B-V)=0.01 from $BVI$ imaging obtained with the Kiso Schmidt telescope.
 
Recent age determinations for the cluster, which employ stellar models including a moderate level of convective core overshoot, are at a reasonable level of agreement with one and other. For example, Meynet, Mermilliod \& Maeder (1993) conclude that the cluster is $\tau$=240Myrs, while Harris et al. (1993) determine an age of $\tau$=200Myrs. Kharchenko et al. (2005) estimate the age of NGC2287 to be $\tau$=280Myrs, while Sharma et al. (2006) favour $\tau$=250Myrs. Thus the weight of evidence favours the age of NGC2287 to lie within the range $\tau$= 243$\pm$40Myrs. However, metalicity estimates for NGC2287 are somewhat scattered. In their catalogue of spectroscopic stellar abundance determinations, Cayrel de Strobel et al. (1992) quote values of [Fe/H]=0.00 and [Fe/H]=-0.25 for the cluster members HD49091 and HD49068 respectively. Cameron (1985) used $UBV$ photometry of 14 members to derive [Fe/H]=+0.065 while from Str\"omgren photometry of F star members, Nissen (1988) determined [Fe/H]=-0.10. Collectively these results point towards the metalicity of NGC2287 being close to the solar value, perhaps marginally less.

The youthful age, the relative proximity and the low line of sight reddening of NGC2287 make it a particularly suitable target for the study of the IFMR, as was recognised several decades ago. Indeed, Romanishin \& Angel (1980) undertook a search of the cluster using photographic plates and this led to the identification of five candidate white dwarf members. Follow-up spectroscopic observations obtained with the 3.6m ESO telescope and the Image Dissector Scanner confirmed at least three of these objects to be white dwarfs, two with $T$$_{\rm eff}$$\approx$25000K and one with $T$$_{\rm eff}$$\approx$13000K (Koester \& Reimers 1981). It was concluded by these authors that the two hotter stars were likely to be cluster members while the cooler white dwarf was probably a foreground object. However, as the quality of the existing spectral data is rather poor, no detailed analyses of the Balmer line profiles in the spectral energy distributions of these stars has ever been undertaken. Thus robust estimates of the effective temperatures and surface gravities of these objects are unavailable to confirm or otherwise these conclusions and to allow them to be fully exploited in the context of the IFMR.  

\section[]{Observations and data analysis}

\subsection{FORS1 spectroscopy of white dwarf candidate members of NGC3532 and NGC2287}

Low resolution, high signal-to-noise optical spectroscopy of the nine white dwarf candidate members of 
the clusters NGC3532 and NGC2287 was obtained in service mode with the European Southern Observatory (ESO) 
Very Large Telescope and the Focal Reducer and low dispersion Spectrograph (FORS1) within the periods  
2007/04/24-27 and 2007/10/06-11/21. A full description of the FORS1 instrument may be found on the ESO 
webpages\footnote{http://www.eso.org/instruments/fors1/}
As these targets are comparatively bright, we specified fairly relaxed constraints on the sky conditions and 
thus the observations were generally undertaken in poorer seeing and/or with some cloud present. All data were
acquired using the 2$\times$2 binning mode of the E2V CCD, the 600B+12 grism and a 1.6" slit which gives a 
nominal resolution of $\lambda$/$\Delta$$\lambda$$\sim$500. Flat and arc exposures were obtained within a few hours
of the acquisition of each of the science frames.

The CCD data were debiased and flat fielded using the IRAF procedure CCDPROC. Cosmic ray hits were removed 
using the routine LACOS SPEC (van Dokkum 2001). Subsequently the spectra were extracted using the APEXTRACT 
package and wavelength calibrated by comparison with the He+HgCd arc spectra. Remaining instrument signature 
was removed using a spectrum of the featureless DC white dwarf WD0000+345 obtained with 
an identical set-up during this programme.

\subsection[]{The model atmosphere calculations}

We have used recent versions of the plane-parallel, hydrostatic, non-local thermodynamic equilibrium (non-LTE)
atmosphere and spectral synthesis codes TLUSTY (v200; Hubeny 1988, Hubeny \& Lanz 1995) and SYNSPEC (v48; Hubeny, 
I. and Lanz, T. 2001, http://nova.astro.umd.edu/) to generate a grid of pure-H synthetic spectra covering the 
$T$$_{\rm eff}$ and surface gravity ranges 14000-35000K and log $g$=7.25-8.75 respectively. We have employed a model 
H atom incorporating the 8 lowest energy levels and one superlevel extending from n=9 to n=80, where the dissolution
of the high lying levels was treated by means of the occupation probability formalism of Hummer \& Mihalas (1988),
generalised to the non-LTE situation by Hubeny, Hummer \& Lanz (1994). The calculations assumed radiative equilibrium 
and included the bound-free and free-free opacities of the H$^{-}$ ion and incorporated a full treatment for
the blanketing effects of HI lines and the Lyman $-\alpha$, $-\beta$ and $-\gamma$ satellite opacities as computed 
by N. Allard (e.g. Allard et al. 2004). During the calculation of the model structure the hydrogen line broadening
was addressed in the following manner: the broadening by heavy perturbers (protons and hydrogen atoms) and electrons
was treated using Allard's data (including the quasi-molecular opacity) and an approximate Stark profile (Hubeny, 
Hummer \& Lanz 1994) respectively. In the spectral synthesis step detailed profiles for the Balmer lines were calculated
from the Stark broadening tables of Lemke (1997).

\subsection[]{Determination of the effective temperatures and surface gravities}

As is our previous work (e.g. Dobbie et al. 2006a), comparison between the models and the data is undertaken using the 
spectral fitting program XSPEC (Shafer et al. 1991). In the present analysis all lines from H-$\beta$ to H-8 are included
in the fitting process. XSPEC works by folding a model through the instrument response before comparing the result to the 
data by means of a $\chi^{2}-$statistic. The best fit model representation of the data is found by incrementing free 
grid parameters in small steps, linearly interpolating between points in the grid, until the value of $\chi^{2}$ is 
minimised. Errors in the $T$$_{\rm eff}$s and log $g$ s are calculated  by stepping the parameter in question away from 
its optimum value and redetermining minimum $\chi^{2}$ until the difference between this and the true minimum $\chi^{2}$ 
corresponds to $1\sigma$ for a given number of free model parameters (e.g. Lampton et al. 1976). The results of our 
fitting procedure are given in Table 1 and shown overplotted on the data in Figure 1.  It should be noted that the 
parameter errors quoted here are formal $1\sigma$ fit errors and undoubtedly underestimate the true uncertainties.
In our subsequent analysis we assume more realistic levels of uncertainty of 2.3\% and 0.07dex in effective temperature 
and surface gravity respectively (e.g. Napiwotzki, Green \& Saffer 1999).

\subsection[]{V band CCD photometry of the white dwarfs}

$V$ band CCD imaging of the nine white dwarfs candidate members of NGC2287 and NGC3532 was obtained on the nights of 2008/03/06
 and 2008/03/07 with the Australia National University's 40'' telescope and the Wide Field Imager (WFI) located at Siding Spring 
Observatory. Conditions on both nights were comparatively good with photometric skies and seeing as measured from the images of 
$\sim$1.5''. The WFI consists of a mosaic of 8 MIT Lincoln Labs 4096$\times$2048 pixel CCDs which covers an area of 
52'$\times$52' in each pointing. Throughout our run, however, CCD6 was non-functional. All data were reduced using the Cambridge 
Astronomical Survey Unit CCD reduction toolkit (Irwin \& Lewis 2001) to follow standard procedures, namely subtraction 
of the bias, flat-fielding, astrometric calibration and stacking. Aperture photometry was performed on the stacked images using 
a circular window with a diameter of 1.5$\times$ the full width half maximum of the mean point spread function. The Landolt fields 
SA98 and SA104 (Landolt 1992) were observed a number of times during the latter night so that instrumental magnitudes could be 
transformed onto the standard $V$ Johnson system, 

\begin{eqnarray}
m_{V}= -2.5 \log (ADU/t_{exp}) + K_{0} + K_{1}X + K_{2}(B-V) \\
\nonumber + K_{3}X(B-V)
\end{eqnarray}

\noindent where $ADU$ is a measure of the total counts from the source, $t_{exp}$ the exposure time and $X$ the airmass. The coefficients
and their respective errors were determined to have the values shown in Table 2. The $B-V$ colour of each white dwarfs was estimated from 
the known effective temperature and surface gravity using the synthetic photometry of Bergeron et al. (1995) as updated by Holberg \& 
Bergeron (2006). Our estimates of the $V$ magnitudes of the nine white dwarfs are listed in the final column of Table 1. 

\section{Discussion}

\subsection{Membership status of the nine white dwarfs}

We have used the estimates of the effective temperatures and surface gravities of the white dwarfs shown in Table 1 and 
the model grids of Bergeron et al. (1995), as revised by Holberg \& Bergeron (2006), to derive absolute 
magnitudes (M$_{V}$; see Table 1). Subsequently, we have determined the distance modulii of the nine white dwarfs, neglecting  
extinction which is believed to be low A$_{V}$$\simless$0.12 along the lines of sight to NGC3532 and NGC2287. These are plotted (solid 
bars) along with a number of distance estimates available in the literature for each of these clusters (dash-dotted 
lines) in Figure 2. It is clear from Figure 2 that NGC3532-6 and NGC3532-8 lie beyond NGC3532 and thus are most probably field 
objects. Moreover, -3 appears to lie to the foreground of NGC2287 as concluded by Koester \& Reimers (1981) and is also probably a 
field white dwarf. The remaining six objects have distance modulii which argue strongly that they are members of NGC3532 or NGC2287. 
These stars are suitable for placing constraints on the form of the IFMR.

\begin{table}
\begin{minipage}{80mm}
\begin{center}
\caption{Coefficients determined for Equation 1, the transformation between instrumental magnitudes and Johnson $V$}

\label{wdvag}
\begin{tabular}{cccc}
\hline
$K_{0}$ & $K_{1}$ & $K_{2}$ & $K_{3}$ \\
\hline
22.326$\pm$0.012 & -0.104$\pm$0.009 & 0.213$\pm$0.014 & -0.138$\pm$0.011\\

\hline
\end{tabular}
\end{center}
\end{minipage}
\end{table}~
\begin{figure}
\vspace{160pt}
\includegraphics{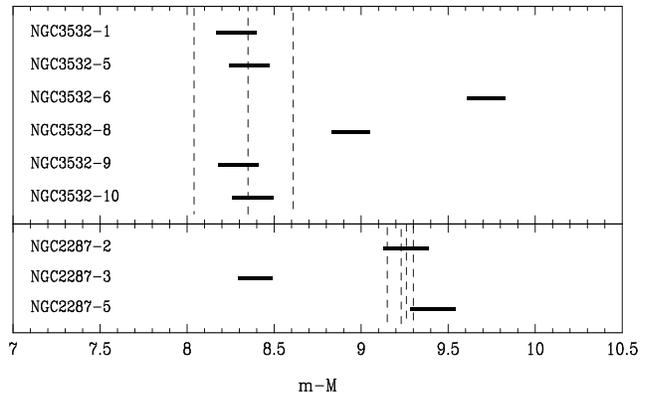}
\caption{The derived distance modulii of the 9 white dwarfs included in this study. The distance modulus of NGC3532 as 
estimated by Meynet, Mermilliod \& Maeder (1993; m-M=8.35), Robichon et al. (1999; m-M=8.04) and Kharchenko et al. (2005; m-M=8.61)
 and the distance modulus of NGC2287 as estimated by Harris (1993; m-M=9.23), Meynet, Mermilliod \& Maeder (1993; m-M=9.15), Kharchenko 
et al. (2005; m-M=9.30) and Sharma et al. (2006; m-M=9.26) are overplotted. The white dwarfs which are non-members are clearly distinguished.}
\end{figure}

\subsection{The IFMR}

\subsubsection{Cluster parameters, initial masses and excluded objects}

\begin{figure*}
\vspace{330pt}
\includegraphics{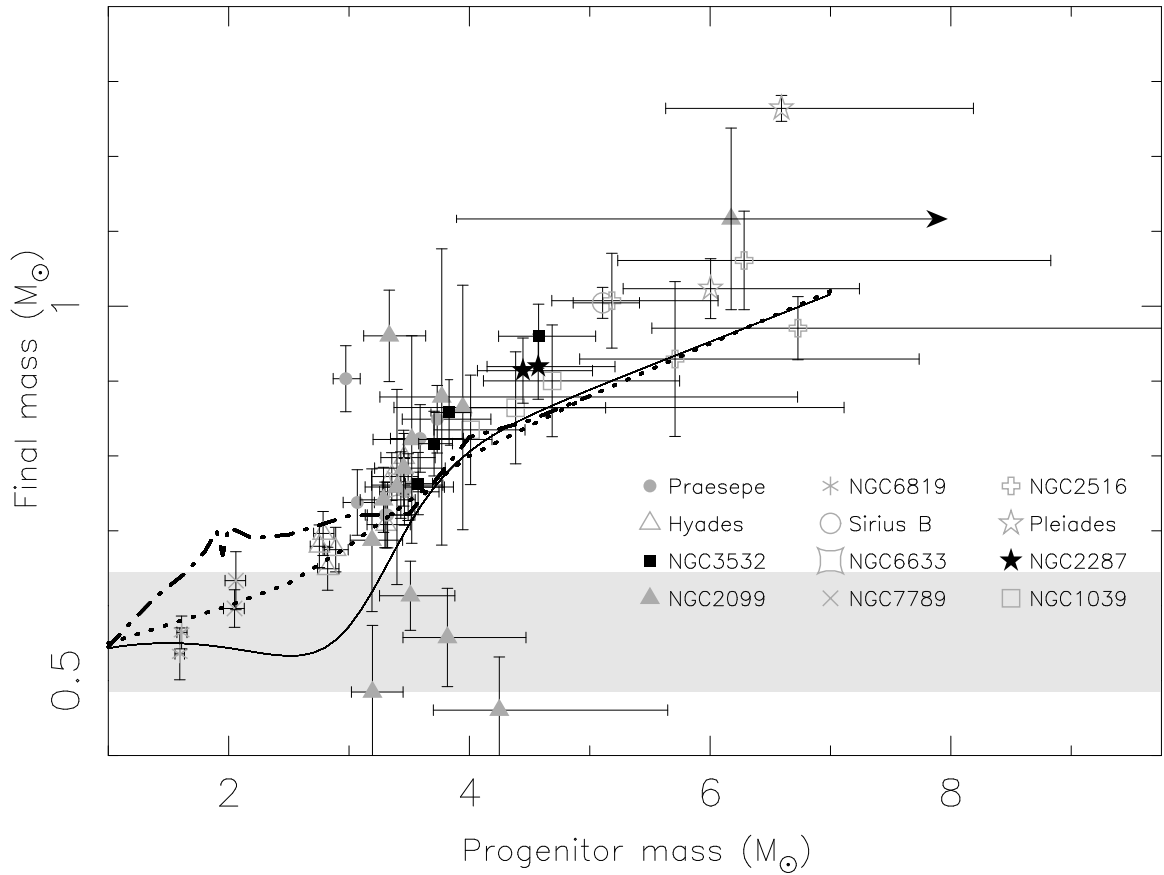}
\caption{The locations of the white dwarf members of NGC3532 and NGC2287 in initial mass-final mass space. Data points from a number 
of other populations with metalicities close to the solar value are also shown (see text for details). The theoretical IFMR of Marigo et al. 
(2007; dot-dashed heavy line), the semi-empirical IFMR of Weidemann (2000; heavy dotted line) and the initial mass - core mass at first thermal 
pulse relation from Karakas, Lattanzio \& Pols (2002; medium solid line) are overplotted. The peak in the field white dwarf mass distribution 
($\pm$1$\sigma$) is represented by the band of grey shading.}
\end{figure*}

The masses and cooling times of the six cluster white dwarfs have been determined using modern evolutionary tracks 
supplied by the Montreal group (e.g. Fontaine, Brassard \& Bergeron 2001). So that this work is generally consistent 
with other recent studies in this area (e.g. Dobbie et al. 2006a, Williams \& Bolte 2007) we have 
adopted the calculations which include a mixed CO core and thick H surface layer structure. The masses and cooling 
times shown in Table 1 have been derived using cubic splines to interpolate between points in this grid. We note that
these mass determinations are relatively insensitive to our choice of core composition and if instead we had adopted 
thin H-layer models these estimates would be systematically lower by only 0.02M$_{\odot}$, which is within our present
level of precision.

The lifetime of the progenitor star of each of these six objects has been derived by subtracting the estimated cooling 
time, shown in Table 1, from the adopted cluster age, where, for the reasons outlined in Section 2, it has been assumed
that $\tau$=243$\pm$40Myrs for NGC2287 and $\tau$=300$\pm25$Myrs for NGC3532. In this calculation, we have taken the errors 
in the cooling times to have magnitudes as shown within the brackets of the relevant column of Table 1. These are based on 
more likely levels of uncertainty in our effective temperature and surface gravity determinations of 2.3\% and 0.07dex 
respectively (e.g. Napiwotzki, Green \& Saffer 1999). Subsequently, we have used cubic splines to interpolate between the 
lifetimes calculated for stars of solar composition\footnote{Based on the Anders \& Grevasse (1989) definition} by Girardi et al.
 (2000) and have constrained the masses of these six 
progenitors to the values shown in the final column of Table 1. We note in this context and in the framework of main sequence 
turn-off based cluster age estimates that current eclipsing binary data are consistent with $\alpha_{\rm OV}$=0.2 (ie. a moderate 
level of) convective core over-shooting across the broad mass range M$\sim$2-30M$_{\odot}$ (Claret 2007). However, we caution
that the errors we quote here in progenitor mass are merely approximations since they have not been determined through a detailed 
statistical analysis (e.g. Salaris et al. 2008). Nevertheless, their magnitudes should provide a guide to the impact of the uncertainties 
in both the white dwarf parameters and the cluster ages which are main sources of error on the final and initial masses respectively 
(Salaris et al. 2008).

The locations of the six white dwarf members of NGC3532 and NGC2287 in initial mass-final mass space are shown plotted in Figure 3,
with data from the extensively studied Sirius system (e.g. Barstow et al. 2005) and a number of other clusters which have metalicities 
that are found to be reasonably close to the solar value (within 30-40\%). While Sirius B is not associated with a particular star cluster,
we include it here since there are relatively few objects in the M$_{\rm init}$$\simgreat$5M$_{\odot}$ regime and the uncertainties on individual 
points here are particularly large. To determine the initial and final masses of the white dwarf members of these additional populations
we have used the same model grids and methodology 
as applied to NGC3532 and NGC2287. For the Pleiades we have assumed an age of $\tau$=125$\pm$25Myrs (e.g. Ferrario et al. 2005) and have 
utilised the white dwarf parameters determined by Dobbie et al. (2006a,2006b). In the cases of NGC6819 (Kalirai et al. 2008), NGC7789 
(Kalirai et al. 2008), the Hyades (Claver et al. 2001), Praesepe (Claver et al. 2001, Casewell et al. 2009), NGC6633 (Williams \& Bolte 2007),
NGC1039 (Rubin et al. 2008) and Sirius (Liebert et al. 2005), we have adopted for the cluster age and the effective temperatures and surface 
gravities of the white dwarf members, the values listed in the relevant referenced work. 

Although Kalirai et al. (2005) adopted an age of $\tau$=650Myrs and a substantially subsolar composition for NGC2099, the results from two recent
spectroscopic studies suggest that the cluster has near solar metalicity.  Marshall et al. (2005) measure [Fe/H]=+0.05$\pm$0.05 from moderate
resolution spectroscopy of eight giant members while Hartman et al. (2008) determine [M/H]=+0.02$\pm$0.04 from high resolution spectroscopy of
candidate members with $T$$_{\rm eff}$$>$4500K. Moreover, during the last fifteen years the bulk of age estimates for NGC2099 obtained using 
theoretical isochrones generated from solar metalicity stellar models which include moderate levels of convective core overshooting, have found 
values within the range $\tau$=450-550Myrs e.g. $\tau$=450Myrs, Mermilliod et al. (1996), $\tau$=520Myrs, Kalirai et al. (2001), $\tau$=450Myrs, 
Kiss et al. (2001) and $\tau$=550Myrs, Hartman et al. (2008). Here we have assumed the cluster has solar metalicity and have adopted the mean of 
the above age determinations, $\tau$$\sim$490$\pm$70Myr, where the error bound has been tuned to envelope the bulk of these estimates. 

Both Ferrario et al. (2005) and Dobbie et al. (2006a) adopted $\tau$=158Myrs for NGC2516, a key cluster for constraining the form of the top end 
of the IFMR. This age was drawn from the work of Sung \& Bessell (2002) and is marginally larger than that adopted by Koester \& Reimers (1996), 
$\tau$=140Myrs which is from the work of Meynet, Mermilliod \& Maeder (1993). Kharchenko et al. (2005) recently derived $\tau$=120Myrs using 2MASS 
PSC data, but this is based on only three cluster stars. Since the work of Ferrario et al. (2005) and Dobbie et al. (2006a), a new detailed 
photometric study of the cluster (Lyra et al. 2006), which used the isochrones of Girardi et al. (2000), has concluded that $\tau$=140Myrs. Looking at 
all these estimates collectively, we conclude that the age of NGC2516 most probably lies within the range 145$\pm$30Myrs.

We have excluded a number of white dwarf candidate members of these clusters from our subsequent analysis for the following reasons: WD0837+218 
is more likely to be a field star than a member of Praesepe (Casewell et al. 2009), WD0836+201 is strongly magnetic and may have a substantially 
different evolutionary history to that of a typical non-magnetic star (e.g. Wickramasinghe \& Ferrario 2000, Tout et al. 2008), WD0836+185, NGC6633 
LAWDS 4 and 7 may be double-degenerate systems as suggested by photometric or radial velocity data, in which case close binary interaction could 
have significantly impacted their evolution, NGC6633 LAWDS 16 is a DB white dwarf for which determinations of effective temperature and surface gravity are
considerably less certain (e.g. Kepler et al. 2007),  NGC2099 WD 6 and 21 do not have spectroscopic surface gravity determinations, NGC2099 WD 15,16 and 17 are found
to be too old at the revised age of their putative parent cluster, NGC1039 LAWDS 20, S1 and 9 have proper motions, as listed in the SuperCOSMOS Sky 
Survey database (Hambly et al. 2001), which are $\sim$3.3$\sigma$, $\sim$4.1$\sigma$ $\sim$2.3$\sigma$ from the mean of their putative parent cluster 
and are thus likely to be field stars (see Figure 4) and NGC7063 LAWDS 1\footnote{The proper motion is listed in the SuperCOSMOS Sky Survey database
 as $\mu_{\alpha}\cos\delta$$\approx$-80$\pm$307 mas yr$^{-1}$ and $\mu_{\delta}$$\approx$-60$\pm$302mas yr$^{-1}$. These huge uncertainties may 
indicate that there was a problem with the measurement.}, with M$\simless$0.4M$_{\odot}$ (Williams \& Bolte 2007), has a proper motion as listed in
the USNO-B1.0 catalogue of $\mu_{\alpha}\cos\delta$=-2$\pm$4 mas yr$^{-1}$ and $\mu_{\delta}$=-22$\pm$8 mas yr$^{-1}$  and as measured by us from 
blue Palomar Sky Survey plates (O269, 1951/07/05 and SJ04686, 1992/07/26) of $\mu_{\alpha}\cos\delta$=2.1$\pm$5.9 mas yr$^{-1}$ and $\mu_{\delta}
$=-19.0$\pm$5.6 mas yr$^{-1}$,  which argues it is more likely to be a field star than a cluster member ($\mu_{\alpha}\cos\delta$=1.24$\pm$0.41 mas
yr$^{-1}$ and $\mu_{\delta}$=-2.83mas yr$^{-1}$; Kharchenko et al. 2005). 

\begin{figure}
\vspace{240pt}
\includegraphics{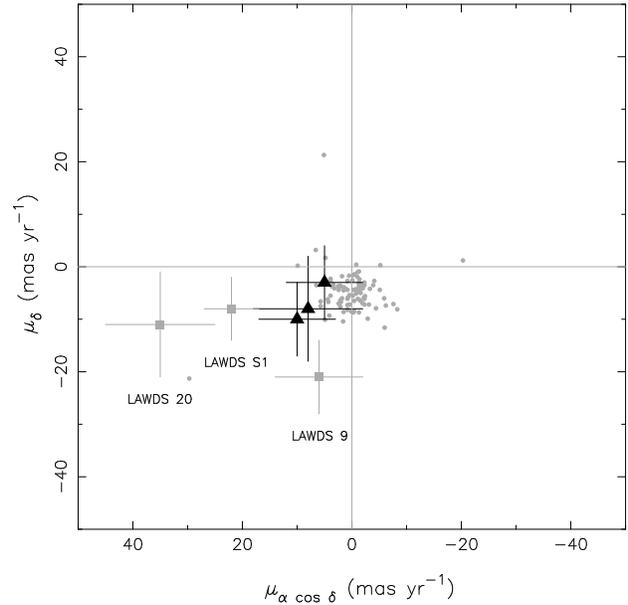}
\caption{SuperCOSMOS proper motion measurements for white dwarf candidate members of NGC1039 recently identified by Rubin et
 al. (2008). The motions of the three massive white dwarfs (filled triangles) are consistent with those of the cluster members
 colour selected by Irwin et al. (2006; small filled circles). The white dwarfs with masses close to 0.6M$_{\odot}$ are also 
shown (filled squares). The proper motions of LAWDS 20 and LAWDS S1 are inconsistent with membership of NGC1039, while the proper motion of LAWDS 9 indicates
it too is more likely to be a field star.}
\end{figure}

\subsubsection{A new look at the form of the IFMR}

What is immediately apparent from Figure 3 is that the white dwarfs of NGC3532 cannot now substantiate any claim that there is significant 
scatter in the IFMR as could their location in the plots of some other recent studies of the IFMR (e.g. Ferrario et al. 2005, Catalan et al. 
2008). The NGC3532 data points in these investigations are based on the older spectroscopy of Reimers \& Koester (1989) and Koester \& Reimers (1993)
which is of lower quality than the data presented here. At the revised cluster age and metalicity, the locations of the majority of the NGC2099 white 
dwarfs now appear to be entirely consistent with those of the Hyades, Praesepe, NGC3532 and NGC6633 clusters. There are only six substantially deviant 
data points amongst the forty-nine shown in Figure 3. It is noteworthy that it is the white dwarf population of the most distant cluster in this sample,
NGC2099, where proper motions are lacking and distance determinations are most unreliable, which presents the highest proportion of outliers, 36\% or 
five out of these 
six. Kalirai et al. (2005) estimate a contamination level of $\sim$25\%, equating to $\sim$6 interlopers in their total spectroscopically observed sample 
of 24 objects, so it seems feasible that at least some proportion of these deviant points are simply field stars. Indeed, the masses of four of these 
objects reside within the peak in the field white dwarf mass distribution, (M=0.565M$_{\odot}$, $\sigma$=0.08M$_{\odot}$, Liebert, Bergeron \& Holberg 2005), 
shown by the grey shaded region. Proper motion measurements for white dwarf candidate members of NGC1039 (Rubin et al. 2008) which appeared to occupy 
this part of initial mass-final mass space (Rubin et al. 2008) confirm that two stars (NGC1039 LAWDS 20 and S1) and argue strongly that a third (NGC1039
LAWDS 9), are simply field objects (see Figure 4). Moreover, with our improved photometry and spectroscopy of the NGC3532 white dwarfs we have also 
demonstrated here that the two candidate members which would otherwise reside in this region of initial mass-final mass space are field stars.

Another possibility that cannot be discounted is that some of these low-lying stars have formed through a close binary evolutionary channel, most probably
where the post-main sequence phases have been terminated prematurely by the loss of envelope mass through the formation of a common envelope (Willems 
\& Kolb 2004). The modelling of Iben \& Tutukov (1985) demonstrates that a M=4M$_{\odot}$ star which experiences Roche Lobe overflow while ascending
the Red Giant Branch can result in the production of a M$_{\rm final}$=0.523M$_{\odot}$ white dwarf. It would be informative, although difficult in 
practice, to obtain near-IR photometry to search for evidence of the presence of cool low mass companions to these NGC2099 white dwarfs once they had 
been shown to have proper motion consistent with the cluster. It has also been suggested a number of times that binarity could have played a role in 
the distinctive location of LB5893 above the bulk of stars in initial mass-final mass space (e.g. Claver et al. 2001, Casewell et al. 2009). Praesepe is
known to harbour a number of blue straggler stars (e.g. 40 Cancri and Epsilon Cancri; Ahumada \& Lapasset 2007). A proportion of these likely form as a 
consequence of mass-transfer between and the eventual coalescense of the components of primordial binaries (e.g. Lombardi et al. 2002). In terms of 
evolution, these stars appear to be retarded with respect to their parent population. Thus it seems plausible that the progenitor of LB5893 was a blue 
straggler and thus the whole evolution of this star has been somewhat delayed with respect to the general Praesepe population. NGC2099 WD11, which also 
sits above the bulk of stars in initial mass-final mass space, might have a similar evolutionary history. 

We find that the current crop of solar metalicity open cluster white dwarfs appears to offer no decisive evidence that non-magnetic stars which have 
effectively evolved in isolation, experience strong differential mass loss, at least within the initial mass range explored here. A number of recent studies 
have shown that a simple linear function, M$_{\rm final}$=M$_{\rm init}$ m + c, is a reasonable approximation to the semi-empirical IFMR. For example, based 
on data from the Sirius binary system and seven open clusters, including the metal poor NGC2168, Ferrario et al. (2005), determined best fit parameters of
m=0.10038$\pm$0.00518 and c=0.43443$\pm$0.01467 over the range 2.5M$_{\odot}$$\simless$M$_{\rm init}$$\simless$6.5M$_{\odot}$. With the addition of data from three more
open clusters, Kalirai et al. (2008) obtained parameters of m=0.109$\pm$0.007 and c=0.394$\pm$0.025 over the range 1.16M$_{\odot}$$\simless$M$_{\rm init}$$\simless$6.5M$
_{\odot}$. However, there is some evidence within the data shown in Figure 3 that the IFMR is somewhat steeper in the range 3M$_{\odot}$$\simless$M$_{\rm init}$$
\simless$4M$_{\odot}$ than for initial masses immediately lower and higher than this. For example, taking the four lowest mass Hyades stars and the white dwarfs 
of NGC7789 and NGC6819, spanning the range 1.6M$_{\odot}$$\simless$M$_{\rm init}$$\simless$3.0M$_{\odot}$, we estimate $\Delta$M$_{\rm final}$/$\Delta$M$_{\rm init}$=0.0981$
\pm$0.0301. For the white dwarfs lying within the range 3M$_{\odot}$$\simless$M$_{\rm init}$$\simless$4M$_{\odot}$, excluding the outliers we determine $\Delta$M$
_{\rm final}$/$\Delta$M$_{\rm init}$=0.2279$^{+0.1151}_{-0.0707}$. Over the initial mass regime 3.8M$_{\odot}$$\simless$M$_{\rm init}$$\simless$5M$_{\odot}$ we estimate 
$\Delta$M$_{\rm final}$/$\Delta$M$_{\rm init}$=0.1231$^{+0.0548}_{-0.0624}$, where for the NGC1039 white dwarfs alone we find $\Delta$M$_{\rm final}$/$\Delta$M$
_{\rm init}$$\le$0.1269. We note that Ferrario et al. (2005) have demonstrated that population synthesis models which adopt an IFMR which has a feature of this 
nature can reproduce the relatively sharp drop observed at the high mass end of the main peak in the mass distribution of white dwarfs (e.g. Kepler et al. 2007, Liebert, 
Bergeron \& Holberg 2005, Marsh et al. 1997). 

The trends outlined by the bulk of the data in Figure 3 bear some resemblance to those of the initial mass-core mass at first thermal pulse relation (thin 
solid line). This theoretical track is likewise relatively flatter at M$_{\rm init}$$\simless$3M$_{\odot}$ since stars with M$_{\rm init}$$\simless$2.3M$_{\odot}$ 
develop comparable degenerate He cores after core-H exhaustion (e.g. Wagenhuber \& Groenewegen 1998). It also becomes notably steeper for 3M$_{\odot}$$\simless$M$_{\rm init}$$
\simless$4M$_{\odot}$, as a result of the sensitivity of the mass of the H-exhausted core to the initial mass (e.g. Becker \& Iben 1979). The slope of this relation 
then decreases at M$_{\rm init}$$\simgreat$4M$_{\odot}$, since here the He-burning shell in the early-AGB phase is predicted to be sufficiently potent to power a 
(second) dredge-up event which reduces the mass of the H-depleted core (Becker \& Iben 1979). The reasons that the IFMR should follow rather closely the initial mass-core
mass at first thermal pulse relation, more especially for M$_{\rm init}$$\simgreat$3M$_{\odot}$, have been discussed in depth by Weidemann (2000). His estimate of the IFMR is overplotted on the data in Figure 3 (heavy dotted line). In brief, during the interpulse period when the He burning shell is inactive, the H-exhausted core
increases in mass. However, the re-ignition of the He-shell in a flash, drives intershell convection which for a short period extends into the CO core, mixing C-rich material
up into this zone. Subsequently, the H-rich convective envelope extends down into the intershell region, dredging processed elements up to the stellar surface and effectively 
reducing the mass of the H-deficient core. If dredge-up is particularly efficient, that is, the mass of material mixed into the predominantly H-envelope is comparable to the 
increase in the mass of the core, $\lambda$$\approx$1, then the core does not grow appreciably while the star evolves on the thermally pulsing AGB. Detailed modelling of AGB
evolution indicates that the maximum $\lambda$ value attained during the TP-AGB is a strong function of initial mass, with M$_{\rm init}$$\simgreat$3M$_{\odot}$ reaching large
$\lambda$ after only a few thermal pulses (e.g. Herwig 2000, Karakas, Lattanzio \& Pols 2002). Indeed, a recent theoretical IFMR for solar metalicity (Marigo \& Girardi 2007; dot-dashed heavy line) reproduces the steepening in the range 3$\simless$M$_{\rm init}$$\simless$4M$_{\odot}$ and the decrease of the slope at M$_{\rm init}$$\simgreat$4M$_{\odot}$. However, the data in M$_{\rm init}$$\simgreat$3.3M$_{\odot}$ regime appear to sit slightly above (a few hundreths of a solar mass) both this and the initial mass-core mass at first thermal pulse relation which may indicate that third dredge-up may not be quite as efficient  here as assumed in the Padova models. Nevertheless, the similarities between the forms of the theoretical relations and the trends delineated by the bulk of white dwarfs from solar metalicity open clusters lend some assurance to the results of modern stellar evolutionary calculations. 

\section{Summary}

We have obtained high signal-to-noise low resolution optical spectroscopy of the nine candidate white dwarfs members of NGC3532 and NGC2287 with FORS1 
and the VLT. The analysis of these data and of new $V$ band photometry indicate that only six of these objects are probably members of the clusters. 
These six objects, in particular the four members of NGC3532, do not substantiate any claim that there is substantial scatter in the IFMR. While a simple 
linear fit to these data could still be deemed acceptable, there are now clear hints that the IFMR is steeper in the initial mass range 3M$_{\odot}$$\simless$M$_{\rm init}$$
\simless$4M$_{\odot}$ than at progenitor masses immediately lower and higher than this. This result is consistent with the predictions of stellar evolutionary 
models. Moreover, it can help explain the relatively sharp drop in the number density of white dwarfs on the high mass side of the main peak in the white dwarf
mass distribution. Unfortunately, the IFMR remains rather poorly constrained at M$_{\rm init}$$\simgreat$5M$_{\odot}$, where there is particular interest in its 
form. Additional white dwarfs and improved spectroscopy on exisiting data points are urgently required in this progenitor mass range.

\section*{Acknowledgments}

MBU and RN are supported by STFC advanced fellowships. We thank Paola Marigo for forwarding to us in tabular form the latest 
Padova theoretical IFMR. We thank the referee, Andre Maeder, for a prompt and helpful report.

\bsp

\label{lastpage}

\end{document}